\input amstex
\documentstyle{amsppt}
\define\a{\alpha}
\redefine\b{\beta}
\redefine\d{\delta}
\define\w{\omega}
\define\nn{\medpagebreak}
\define\n{\smallpagebreak}
\define\nnn{\bigpagebreak}
\define\s{\sigma}
\define\g{\gamma}
\define\ar{\longrightarrow}
\redefine\e{\epsilon}
\document
\topmatter
\title
Computations on nondeterministic cellular automata
\endtitle
\author
Y.I.Ozhigov
\endauthor
\endtopmatter
\nn

 Department of Mathematics, Moscow State
                Technological University "Stankin", Vadkovsky per.
3a,
                 101472, Moscow, Russia

{\it e-mail:} y\@oz.msk.ru
\n

\specialhead Abstract \endspecialhead
\n

The work is concerned with the trade-offs between the dimension
and the time and space complexity of computations  on
nondeterministic
cellular automata. We assume that the space complexity is the diameter
of area in space involved in computation.

It is proved, that 

1). Every NCA $\Cal A$ of dimension $r$, computing a predicate $P$ with 
time complexity $T(n)$ and space complexity $S(n)$ can be simulated by 
$r$-dimensional NCA with time and space complexity $O(T^{\frac{1}{r+1}} 
S^{\frac{r}{r+1}} )$ and by $r+1$-dimensional NCA with time and space 
complexity $O(T^{1/2} +S)$, where $T$ and $S$ are functions, constructible 
in time.

2) For any predicate $P$ and integer $r>1$ if $\Cal A$ is a fastest 
$r$-dimensional NCA computing $P$ with time complexity $T(n)$ and space 
complexity $S(n)$
 , then $T= O(S)$.

3). If $T_{r,P}$ is time complexity of a fastest $r$-dimensional NCA computing 
predicate $P$ then 
$$
\aligned
T_{r+1,P} &=O((T_{r,P})^{1-r/(r+1)^2} ),\\
T_{r-1,P} &=O((T_{r,P})^{1+2/r} ).
\endaligned
$$

Similar problems for deterministic CA are discussed.

\newpage
\define\NC{\operatorname{NC}}
\define\Mark{\operatorname{Mark}}
\define\con{\operatorname{con}}
\define\out{\operatorname{out}}
\define\In{\operatorname{in}}
\define\Diam{\operatorname{Diam}}
\define\env{\operatorname{env}}

\redefine\succ{\operatorname{succ}}
\define\res{\operatorname{res}}
\redefine\Im{\operatorname{Im}}
\specialhead
\ \ \ \ \ \ \ \ \ \ \ \ 1.Introduction \endspecialhead
\nn

It is well known that nondeterministic computations are more
powerful than deterministic ones. The interrelation between deterministic 
and nondeterministic time complexity of computations was established 
by S.Cook in the work \cite{2} where he showed the existance of NP-complete
problems. However, it is still unknown can nondeterministic computations 
be fulfilled on physical devices or not. 
In this paper we show how the complexity of computations on a nondeterministic
device depend on its dimension. Note that the similar problem for 
deterministic computers is open (look at the section 6).

 Cellular automata (CA) provide a convenient framework for 
studies on this problem. A cellular automaton is a dynamical system with local
interactions operating in discrete space and time, and simultaneously 
CA may be used as a general model of computational device.
CA were introduced by S.Ulam in the work \cite{7} and J. von Neumann in the 
work \cite{5} and 
since then various problems pertaining to CA were treated in a great many 
works (look, for example, at \cite{1},\cite{3},\cite{6},\cite{9},\cite{11},
\cite{12}).

Let $r\geq 1$ be an integer, $\Bbb Z ^r$ be the space, $\w$ be a finite 
alphabet for possible states of any cell $\bar i\in\Bbb Z^r$.
 Cellular automaton of dimension $r$ in alphabet $\w$ is a function of 
the form: $\Cal A :\w ^{2r+1} \ar \w$ .

A cellular automaton determines the special class of evolutions
in $\Bbb Z ^r$ so that the states of all cells 
$ \bar i\in\Bbb Z^r$ evolve synchronously in discrete time steps according to 
the states of their nearest neighbours.

 It must be mentioned, that a more general approach can be of interest for
applications, where the function $\Cal A$ depends on $\bar i$ or $t$.
 It is not our intention to regard such possibilities here.

A configuration is an ensemble of states of all cells at 
some instant of time. 
It is apparent that an evolution of CA is uniquely determined by
the initial configuration.

If we consider a multifunction instead of $\Cal A$ , we obtain
the definition of a nondeterministic cellular automaton (NCA).
Generally speaking, evolutions of NCA are not uniquely determined 
by initial configurations. A behaviour of NCA may be described by a state 
transition network (look at \cite{4}). It is a graph, each of whose nodes 
represents some configuration. Directed arcs join the nodes to represent the 
transition between configurations. All nodes have out-degree one iff the 
cellular automaton is deterministic. Moreover, if the cellular automaton at 
hand is in fact nondeterministic and we consider the configurations in 
unlimited space $\Bbb Z^r$, then out-degrees of some nodes in the state 
transition network will be infinitely large.

The difference between one and high dimensional CA 
has emerged from the solution of the predecessor existence problem (PEP) for CA.
It is the problem of existence of a predecessor for the given configuration.
 S.Wolfram showed in article \cite{8} that PEP 
is decidable for one dimensional CA, and T.Taku in article \cite{10} 
showed that PEP is undecidable for some CA of
dimension $r$ where $r=2,3,\dots$.

 The computational equivalence of CA and Turing Machines is a well known fact 
(look at \cite{2},\cite{11}). 
The time complexity $\Cal T (n)$ and the space complexity $\Cal S (n)$ can be 
defined routinely for any CA $\Cal A$.

 This brings up the question:
given an arbitrary predicate $\Cal P$, how does the minimal complexity of $r$ 
dimensional CA, computing $\Cal P$, depend on $r$?

More precisely, if some predicate $\Cal P$ is computable on CA of dimension 
$r$ with time  complexity $\Cal T (n)>O(n)$ is it possible to compute $\Cal P$ 
substantially faster on CA of dimension $r' >r$ ?

This is called TCD-problem. This problem is open for CA.

A different situation arises with TCD-problem for NCA.
For example, it is found that if some predicate $\Cal P$ 
can be computed on NCA with time complexity $T(n)=O(n^\a )$ and space 
complexity $O(n^{\a /2})$ then the increase of dimension 
by one unit allows to compute $\Cal P$ in time $O(n^{\a /2})$.
 Moreover, given $\b >0$, the  time complexity $O(n^{\b} )$ can be attained 
for the computation of such predicate if we increase the dimension of NCA to 
a suitable value. A similar result takes place also for faster increasing 
functions $T(n)$. 
It means that multi-dimensional NCA will become
 the faster instrument for 
computation as the dimension increases. 

We proceed with the exact definitions. All constants are assumed to depend on 
the dimension $r$.

\nnn
\specialhead \ \ \ \ \ \ \ \ \ \ \ \ 2. The main definitions and results 
\endspecialhead
\nnn

Let $\w =\{ c_0 ,\dots ,c_k \}$ be an alphabet for possible states of cells. 
Let $t$ takes the values from the set $\Bbb N =\{ 0,1,\dots \}$.

An {\it evolution} in $\Bbb Z ^r$ is a function of the form 
$a:\ \Bbb N\times\Bbb Z ^r \ar\w$.
A  {\it configuration} is a function of the form $a^{(t)} : \ \ Z^r \ar\w$.
Any evolution $a$ may be displayed as a sequence 
$$
a^{(0)} ,a^{(1)} ,\dots 
\tag 1 
$$
of configurations at the instants of time $t=0,1,\dots $, 
where $a^{(t)} (\bar i )=a(t,\bar i )$.
$j$th componenet of $i \in \Bbb Z ^r$ will be denoted by $i_j$. If 
${\bar i}=(i_1 ,i_2 ,\dots ,i_r )\in Z^r$,
we shall write $a(t,i_1 ,\dots ,i_r )$ instead of $a(t,{\bar i})$.

The following notations are fixed for the cells $\bar i (j)$ comprising the 
neighborhood of the cell $\bar i$:
$$
\aligned
\bar i (0) &= \bar i ,\\
\bar i (1) &=(i_1 -1 ,i_2 ,\dots ,i_r ),\\
\bar i (2) &=(i_1 +1 ,i_2 ,\dots ,i_r ),\\
\bar i (3) &=(i_1  ,i_2 -1 , i_3 ,\dots ,i_r ),\\
\bar i (4) &=(i_1  ,i_2 +1 , i_3 ,\dots ,i_r ),\\
&\dots  \\
\bar i (2r) &=(i_1  ,i_2 ,\dots ,i_r +1 ).
\endaligned
$$

We put : $a_j (t,\bar i )=a(t,\bar i (j))$.

We'll consider only such evolutions $a$ that $\exists C: \ \forall \bar i : 
\Vert\bar i\Vert >C\ \ \
a(t,\bar i )=c_0$, therefore all configurations $a^{(t)}$ will be finite 
objects.

Let $l_r$ be a fixed computable one-to -one mapping 
$l_r :\ \Bbb N\ar\Bbb Z^r$ such that $\Vert l_r (n)\Vert =O(n^{1/r} )$. 
This function represents an embedding 
of one dimensional space into $r$ dimensional with the least norm 
$\Vert l_r (n)\Vert$.

Given an alphabet $\sigma $, the set of all words over $\sigma$ is denoted 
by $\sigma ^*$.
If $B=c_{j_1} c_{j_2} \dots c_{j_s} \in\w ^*$, 
then the initial configuration corresponding to the word $B$ is defined by
$$
a^{(0)} (\bar i )=\left\{\aligned
c_{j_k} \ \ &\text{if} \ 0\leq l_r^{-1} (\bar i)=k\leq s,\\
c_0 \ \ &\text{otherwise,}\endaligned \right.
$$
we denote this configuration by $a_B^{(0)}$.
\n

A {\it nondeterministic cellular automaton} of $r$ dimensions is a function of 
the form
$$
\Cal A : \  \undersetbrace 2r+1 \ \text{times} \to{\w \times \w \times \dots 
\times \w  } \ar 2^{\w} .
$$

Let $\bar b \in \w ^{2r+1} ,\ c\in\Cal A (\bar b )$. Any word of the form
$\bar b \ar c$ is called a command of this automaton. This command is called 
trivial if $\Cal A (\bar b )=\{ c\}$ and $\bar b$ has the form
$(c,u_1 ,\dots ,u_{2r} )$. A set $G$ of commands of $\Cal A$ which contains 
all nontrivial commands is called a {\it program} of $\Cal A$.
The behaviour of $\Cal A$ is defined by its program.

If $\forall  \bar b\in \w ^{2r+1} $ and $\Cal A (\bar b )$ consists of exactly 
one element, then in fact
 $\Cal A$ has the form $\w ^{2r+1} \ar \w$, and we obtain the definition of a 
deterministic cellular automaton.

We assume that $\Cal A (c_0 ,\dots ,c_0 )=c_0$.
This letter $c_0$ plays the role of blank, it is denoted by $0(\w )$.
\n

 An evolution of NCA $\Cal A$ is a sequence $\hat a$ of the form (1)
where $\forall t=0,1,\dots ,\ \ \forall\bar i\in Z^r $
$$
 a(t+1 ,\bar i )\in\Cal A (a_0 (t,\bar i ),a_1 (t,\bar i ),\dots a_{2r} 
(t,\bar i )).
\tag 2$$

It is obvious that in deterministic case $a^{(t)}$ depends on $\Cal A ,t ,
a^{(0)}$,
and in nondeterministic case, in addition, on the choice of elements (2).

The set of all values of a function $f$ is denoted by $\Im f$.

Let the alphabet $\w$ be divided into two nonintersecting parts:  
$\w =\w '\cup \w ''$, where
 $\w '$ is the set of main letters, $\w ''$ is the set of auxiliary letters, 
and let $E\subset \w ''$ be the set of end letters,
where $c_{k-1} ,c_k \in E$. We denote ${\w '}^*$ by $\Sigma ,\ c_k$ by 
$\succ (\Cal A )$,
 $E$ by $E(\Cal A )$.
Let  for the evolution (1) of NCA $\Cal A$ $\ \Im\ a^{(0)} \subseteq \w ' 
\cup\{ c_0 \}$.
Let $\tau (\hat a)$ be the least value of $t$ such that there 
exists one and only one letter $c\in E\cap\Im a^{(t)}$. This letter $c$ 
is denoted by $\res (\hat a ,\Cal A )$ and is called the result of the 
operation of 
$\Cal A$ on the initial configuration
 $a^{(0)}$ in evolution $\hat a$. 
 A configuration $\hat a_{\tau (\hat a )}$ is called a resulting configuration 
for $a^{(0)}$.
\n

In general terms, the result of the operation of $\Cal A$ is defined uniquely 
only in the deterministic case.
The set of all results in evolutions which begin with $a^{(0)}$ is denoted 
by $\Cal A [a^{(0)} ]$.

A predicate $P$ on the set $\Sigma$ is an arbitrary subset of $\Sigma$.

 NCA $\Cal A$ computes a predicate $P$ iff $\forall B\in \Sigma$
$$
\left\{\aligned  \succ (\Cal A )\in \Cal A [a_B ^{(0)}] ,\ \ &\text{if} 
\ B\in P ,\\
 \succ (\Cal A )\notin \Cal A [a_B ^{(0)}] , \ \ &\text{if} \ B\notin P. 
\endaligned\right.
$$
It's obvious that such a predicate $P$ is defined uniquely for $\Cal A$ if it
exists.
 We denote this predicate by $P_{\Cal A}$.
A cell $\bar i \in Z^r$ is called accessible in evolution $\hat a$ iff 
$\exists t'\leq \tau (\hat a) :\ a(t',\bar i )\neq c_0 $.

The diameter of the set of all accessible cells is denoted by $D (\hat a)$.
 
Given $B\in P$, the least value of $\tau (\hat a )$
from all evolutions $\hat a$ , where $a^{(0)} =a^{(0)}_B ,\ \res (\hat a ,
\Cal A )=\succ (\Cal A )$ is denoted by $\tau _{\Cal A} (B)$.

Let $D_{\Cal A} (B)$ denotes the least value of $D(\hat a )$ from all 
evolutions
$\hat a$ where $a^{(0)} =a^{(0)}_B ,\ \res (\hat a ,
\Cal A )=\succ (\Cal A ),\ \tau (\hat a )=\tau _\Cal A (B)$.

\nn

The {\it time complexity} of NCA $\Cal A$ is the function $\Cal T_{\Cal A} : 
\ \Bbb N\ar\Bbb N$ defined by
$$
\Cal T _{\Cal A} (n) =max \{\tau _{\Cal A} (B)\ |\ B\in P_{\Cal A} ,
\ |B|\leq n\} ,
$$
where $|B|$ denotes the length of $B$.

\n
The {\it space complexity} of $\Cal A$ is the function $\Cal S_{\Cal A} : 
\ \Bbb N\ar \Bbb N$ defined by
$$
S_{\Cal A} (n) = max \{ D_{\Cal A} (B) \ |\ B\in P_{\Cal A} ,\ |B|\leq n\} ,
$$

\n
Without loss of generality we may anticipate that $\Cal A (c_i ,\dots )=c_i$ 
for all $ c_i \in E$.

 Functions $\Cal T _{\Cal A} (n) $ and $\Cal S _{\Cal A} (n)$ can be very 
complicated, and it's convenient to use their best upper approximations  
$T_{\Cal A},\ S_{\Cal A} \in\Cal K :
 \ \ \Cal T _{\Cal A} (n)=O(T_{\Cal A} (n)) ,\ 
\Cal S _{\Cal A} (n)=O(S_{\Cal A} (n))$ instead of them, where $\Cal K$ is the 
class of functions constructible in time (see below). 
 Given the space $\Bbb Z^r$, a function $f:\ \Bbb N\ar \Bbb N$ is called 
{\it constructible in time} on $r$-dimensional NCA, if there exists a constant 
$c$ and NCA $\Cal A$ with time complexity $\Cal T (n)= c(f(n)+n^{1/r} )$ such 
that for every word $B\in\w^* ,\ |B|=n$ there exists the single resulting 
configuration for $a_B^{(0)}$ which has the form
$$
a_\tau (\bar i )=\left\{\aligned
c_1 ,\ &\text{if}\ \bar i\in\{ 1,2,\dots ,\Cal T (n)/2 \}^r ,\\
c_0 ,\ &\text{in the opposite case.}
\endaligned\right.
$$
It means that $r$-dimensional cube of side $\Cal T(n)/2$ can be isolated in 
time $\Cal T(n)$, where $O(n^{1/r} )$ is the size of necessary domain for 
input word $B$.

For example, the constructibility in time for $n^4$ and $2^{2n}$ is in fact 
proved in the section 4 (group G1), for the functions
$n^\a ,q^{\a n} ,\ q,\a\in\Bbb Q$ and for their combinations with additions, 
multiplications and superpositions the constructibility in time may be proved 
along similar lines.

\n
A pair of functions $(T_\Cal A ,S_\Cal A )$ is called a complexity of NCA 
$\Cal A$. Thus in what follows $T_{\Cal A}, \ S_{\Cal A}$ (or, simply $T,\ S$ 
with or without indices) will be constructible in time.

The class of predicates $P$, computable on NCA of dimension $r$ with 
complexity $(T,S)$ is denoted by $\NC (r,T,S)$.

We'll write $T_1 <O(T)$ instead of the following: $\forall C>0\ \exists N\ 
\forall n\geq
N\ T_1 (n)\leq CT(n)$.

$r$-dimensional NCA, computing predicate $P$ with complexity $(T,S)$ is 
called a {\it fastest} NCA if $P$ can not be computed on $r$-dimensional 
NCA in time $T_1 <O(T)$.

Let $T_{r,P}$ denote the time complexity of a fastest $r$-dimensional NCA 
computing predicate $P$.

Here are the main results of this paper.
\nn

\proclaim {Theorem 1}
$$
\NC (r,T,S)\subseteq \NC (r+1,\sqrt{T} +S ,\sqrt{T} +S).
$$

\endproclaim
\n

\proclaim{Theorem 2} 
$$
\NC (r,T,S)\subseteq \NC (r,T_1 , T_1 ),
$$
where $T_1 =T^{\frac{1}{r+1}} S^{\frac{r}{r+1}}$
\endproclaim 
\n

\proclaim {Theorem 3} Let $\Cal A$ be a fastest NCA of dimension $r$.
Then
$$
T_{\Cal A} (n)=O(S_{\Cal A} (n)) .
\tag 3$$
\endproclaim
\n

\proclaim{Theorem 4}  
$$
\aligned
1). \ T_{r-1,P} &=O((T_{r,P})^{1+2/r} ),\\
2).\ T_{r+1,P} &=O((T_{r,P})^{1-r/(r+1)^2} ).
\endaligned
$$
\endproclaim
\nnn

Now we shall give the outline of the following sections.
 All these results are based on two main methods of speeding up computations:
The method of direct simulation in $r+1$ dimensional space (section 3,
Proof of Theorem 1) and the method of optimization of NCA in the same
space (section 4, Proof of Theorem 2). Theorem 3 will be simply 
derived from Theorem 2. Point 1) of Theorem 4 will be proved by the method
of simulation in $r-1$ dimensional space (method of evolvents, section 5).
Point 2) of Theorem 4 will be proved in two steps: reduction of space
complexity in $r+1$ space and the following optimization. 

Note that the both two method of speeding up require nondeterminism.
\nnn

\specialhead \ \ \ \ \ \ \ \ \ \ \ \ 3. Simulation 
in $r+1$-dimensional space. Direct method \endspecialhead

\subsubhead Proof of Theorem 1 \endsubsubhead
\nn

Let $P=P_{\Cal A} \in \NC (r,T,S)$, $\Cal A$ be a cellular automaton of 
dimension  $r$ with alphabet $\w$ and complexity $(T,S)$, $0(\w )=c_0$.

In this section we'll present the direct method of 
speeding up: we shall construct
 the nondeterministic cellular automaton
NCA1 of dimension $r+1$, which simulates $\Cal A$ in time $O(T^{1/2} +S)$.

The rough idea is that we expand the alphabet of the cellular automaton
$\Cal A$ at hand and use $r+1$st dimension to code $H$ state transitions
of $\Cal A$ into one big $r+1$ dimensional state transition of the new
automaton NCA1 simulating $\Cal A$, where $H=O(\sqrt{T} +S)$.
 The single obstacle which will remain is that the initial configuration for 
NCA1 is not $a_B ^{(0)}$, but the
ascending map, corresponding to input word $B$ (the definition is in the next 
section). This obstacle will be overcome in the last part of this section.
\nn

{\bf Definition} A {\it port} is a list $p$ of the form:
$$
p=(\Mark (p),\con (p),\env _1 (p),\env_2 (p) ,\dots ,\env_{2r+2} (p)) ,
\tag 4
$$
where  $A, B$ are the special new letters,
$\Mark (p)\in \{ A,D\} $ , the other members of list (4)
are arbitrary letters from $\w$, and the functions 
$\con (p) ,\env_j (p) ,\ j=1,\dots ,2r+2$ are
defined by equality (4).
\nn

\head Alphabet of NCA1 \endhead
\nn

Let $\w _0$ be the set of all ports with the exception of 
$(D,c_0 ,\dots ,c_0 )$. Then the alphabet of NCA1 is $\w _1 =\w _0 
\cup\{\emptyset ,b\}$, where $\emptyset ,b$ are new letters, and let 
$0(\w _1 )=(A,c_0 ,\dots ,c_0 )$.

\nn
\head Commands of NCA1 \endhead
\nn

 Since the dimension of NCA1 is $r+1$, the commands of it have the following 
form:
$$
(p_0 ,p_1 ,\dots ,p_{2r+2} )\ar \{ p\} .
$$

Let all $p_q \in \w _1 ,\ q=0,1,\dots ,2r+2$. The set $\{ p\}$  consists of 
all elements
$p\in \w _1$ to be described below. Let $s^+ =\{ 0,2,3,4,\dots ,2r+2\} ,
s^- =\{ 0,1,3,4,5,\dots ,2r+2\}$,
$$
k(j)=\left\{\aligned 
j+1 ,\ &\text{for}\ j\ \text{odd},\\
j-1 ,\ &\text{for}\ j\ \text{even}.
\endaligned\right.
$$
We'll consider separately five different cases.
\n

\subhead Case 1 \endsubhead
Let the following conditions be satisfied:

$$
 \forall j\in s^+ \ \ p_j \in \w _0 ,\ \Mark (p_j )=A,
$$
$$\forall j\in s^+ -\{ 0\} \ \ \con (p_j )=\env _j (p_0 ),\ 
\con (p_0 )=\env _k (p_j ),
$$
where $k=k(j)$ and 
$$
\con (p_2 )\in \Cal A (\con (p_0 ),\con (p_3 ) ,\dots ,\con (p_{2r+2} )).
$$

In this case $p$ is such a port that $\Mark (p)=D$, or $p$ is $p_0$.
\n

\subhead Case 2 \endsubhead
Let the following conditions be satisfied:

$$
 \forall j\in s^- \ \ p_j \in \w _0 ,\ \Mark (p_j )=D \ \text{or}\ p_j =
0(\w _1 ),
$$
$$\forall j\in s^- -\{ 0\} \ \ \con (p_j )=\env _j (p_0 ),\ \con (p_0 )=
\env _k (p_j ),
$$
where $k=k(j)$ and 
$$
\con (p_1 )\in \Cal A (\con (p_0 ),\con (p_3 ) ,\dots ,\con (p_{2r+2} )).
$$

In this case $p$ is such a port that $\Mark (p)=A$, or $p$ is $p_0$.
\n

\subhead Case 3 \endsubhead
Let $p_1 =b,\ \Mark (p_0 )=D$ or $p_2 =b$.

 Then $p$ is $p_0$ or $p$ is obtained from $p_0$ by the following redefinition: we put 
$$
\Mark (p)=\left\{ \aligned A, \ \ &\text{if} \ \Mark (p_0 )=D ,\\
D,\ \ &\text{if} \ \Mark (p_0 )=A .\endaligned\right.
$$
\n

\subhead Case 4 \endsubhead
If $\exists j\in s^+ \cap s^- :\ p_j =b$, we put $p=b$.
\n

\subhead Case 5 \endsubhead
In all other cases we put $p=\emptyset$.
\n

Note that NCA1 is nondeterministic even though $\Cal A$ may be a CA of 
deterministic type,  because in the cases 1 and 2 the choice of $p$ is 
not uniquely defined.
\nn

Let $t$ be time step, $H$ be some positive integer. We suppose that 
$H$ is arbitrary till Lemma 3.

Some peculiar configurations of $NCA1$ are called $H,t$-maps.
Maps will be of two sorts.
\n

{\bf Definition}. An {\it Ascending map} of high $H$ at time step $t$ 
($A,H,t$ -map )
is such a configuration $a^{(t)}$ for NCA1 that $\forall \bar i \in 
\Bbb Z ^{r+1}$

$1^0$ If $i_1 \notin \{ -1,H\}$, then $a^{(t)} (\bar i )\in\w _0$ .
If $a(t,\bar i )\neq 0(\w _1 )$, then $1\leq i_1 \leq H$, and $a(t,-1,i_2 ,
\dots, i_{r+1} )=a(t,H,i_2 ,\dots i_{r+1} )$ $=b$; 

$2^0$ If $\bar i :\ \ 0\leq i_1 <H-1$, then
$$
\con (a(t,i_1 +1 ,i_2 ,i_3 ,\dots ,i_{r+1} ))=\Cal A (\con (a_0 (t,\bar i )),
\con (a_3 (t,\bar i )),\dots ,\con (a_{2r+2} (t,\bar i ))),
$$

$3^0$ If 
$$
s(j)=\left\{\aligned 2j,\ \ &\text{if} \ \e =1, \\
2j-1,\ \ &\text{if} \ \e=-1 ,\endaligned\right.
$$
then 
$\forall j:\ 1\leq j\leq r+1 \ \exists \e\in\{ 1,-1\}\ 
\con (a(t,i_1 ,\dots ,i_j +\e ,\dots ,i_{r+1} ))=
\env _k (a(t,\bar i ))$, if the two parts of this equality exist.

\n
{\bf Definition}. A {\it Descending map} of high $H$ at time step $t$ 
($D,H,t$-map) is such a configuration $a^{(t)}$ for $NCA1$, that $\forall 
\bar i \in \Bbb Z ^{r+1}$

$1^0$ See above.

$2^0$ If $ 0< i_1 <H$, then
$$
\con (a(t,i_1 -1 ,i_2 ,i_3 ,\dots ,i_{r+1} ))=\Cal A (\con (a_0 (t,\bar i )),
\con(a_3 (t,\bar i )),\dots ,\con (a_{2r+2} (t,\bar i )).
$$

$3^0$ See above.

The following proposition relates evolutions of $\Cal A$ to those of NCA1.
\nn

\proclaim{Lemma 1} $\a ^{(0)} , \dots ,\a ^{(\tau )} ,\dots $ is evolution of 
$\Cal A$ iff
there exists an evolution
$$
a^{(0)} ,\dots a^{(t)} ,\dots 
\tag 5$$
of NCA1, where for every $\tau =0,1,\dots $ the following condition $C_\tau$ 
is fulfilled:

$C_\tau$. $\exists t=t(\tau )\in\Bbb N \ \exists q=q(\tau ): \ 0\leq q\leq H-1  \ \forall \bar i\in\Bbb Z^r \ \con (a(t,q,i_1 ,i_2 ,\dots ,i_r ))=\a (\tau ,\bar i )$, where
for $t$ even: $\tau =t(H-1)+q$, and $a^{(t)}$ is $A,H,t$-map,
for $t$ odd: $\tau =(t+1)(H-1) -q$ and $a^{(t)}$ is $D,H,t$-map.
\endproclaim
\nn

\subsubhead Proof \endsubsubhead
\n

1. Necessity. Let $\a$ be an evolution of $\Cal A$. 
An evolution $a$ of NCA1 is called $\tau$-evolution if the condition
$C_\tau$ is fulfilled. At first let us prove by induction on $\tau$ that 
{\it for any $\tau$ there exists $\tau$-evolution}. Basis: $\tau =0$ follows 
from the definition of $A,H,t$-map.
Step: follows from the definition of NCA1. $\square$

Now we introduce the order $\prec$ on $\Bbb N\times \{ 0,1,\dots ,H-1\}$ by 
the following: $(t_1 ,q_1 )\prec (t_2 ,q_2 )$ iff $t_1 \prec  t_2$ or 
($t_1 =t_2$ -  even and $q_1 \prec q_2$) or ($t_1 =t_2$ -  odd and 
$q_2 \prec q_1$). Note that if $\tau _1 <\tau _2$, then $(t(\tau _1 ),
q(\tau _1 ))\preceq (t(\tau _2 ),q (\tau _2 ))$. Consequently, in view of the 
definition of $\tau$-evolution, if $a$ is $\tau _1$-evolution, $d$ is 
$\tau _2$-evolution, then for every pair $(t,q)\preceq (t(\tau _1 ),
q(\tau _1 ))$ we have $\forall \bar i\in\Bbb Z^r \ a(t,q,\bar i )=
d(t,q,\bar i )$. Thus we obtain that there exists the evolution $a$ of NCA1 
which coincides with every $\tau$-evolution on all lists $(t,q,\bar i )$, 
where $(t,q)\preceq (t(\tau ),q(\tau ))$. Necessity is proved.

2. Sufficiency. Follows from the definition of NCA1. Lemma 1 is proved.
\nn

\proclaim{Lemma 2} If $a^{(0)} ,a^{(1)} ,\dots a^{(t)} ,\dots $ is evolution 
of NCA1 and 
$a^{(0)} $ is $H,0$-map, then $a^{(t)}$ is $H,t$-map iff 
$$
\forall \bar i\in \Bbb Z ^{r+1} :\ 0\leq i_1\leq H-1\ \ 
a(t,\bar i )\neq \emptyset .
\tag 6
$$ \endproclaim
\nn

\subsubhead Proof \endsubsubhead
\n

Induction on $t$. Basis: $t=0$. Follows from the condition. Step follows from 
the definition of NCA1 and those of $H,t$-map. Lemma 2 is proved.
\nn

Let $[x]$ denote integral part of $x\in\Bbb R$.
\n

\proclaim{Lemma 3} Let some predicate $P$ be computed on NCA $\Cal A$ with 
complexity $(T,S)$, $|H-[\sqrt{T(n)} ]|<\sqrt{T(n)}/2$ , $B\in \Sigma ,\ 
\a ^{(0)} $ be
 the initial configuration of $\Cal A$, corresponding to $B$, 
$t=2[\sqrt{T(n)} ] ,\ n=|B|$.

If $B\in P$ then for some evolution of NCA1 of the form (5) the condition (6) 
is fulfilled and 
$$
\exists \bar i\in\Bbb Z^{r+1}  : \con (a(t,\bar i ))=\succ (\Cal A )
\tag 7$$
 and if $A\notin P$ then for any evolution of NCA1 of the form (5) with the 
condition (6) 
the following property takes place
$$
\forall \bar i : \con (a(t,\bar i ))\neq \succ (\Cal A ) .
\tag 8$$
 \endproclaim
\nn

\subsubhead Proof \endsubsubhead
\n

Follows immediately from the definition of computation on NCA, Lemma 1 and 
Lemma 2. Lemma 3 is proved.
\nn

 \ \ \ \ \ \ \ \ \ \ \ \ {\bf 4. Auxiliary automata}
\nn

To finish the proof of Theorem 1 we must construct two auxiliary NCA:

1). NCA2 -- which begins to operate from $d^{(0)}$ -- initial configuration, 
corresponding to $B\in \Sigma$ in $\Bbb Z^{r+1}$ , so that for some evolution 
$a$ from Lemma 1 and for some $t'$ $\a ^{(t')} =a^{(0)}$, 
$|H-[\sqrt{T(n)}]|<\sqrt{T(n)} /2,\ n=|B|$.

2) $CA3$ -- deterministic CA, which begins to operate from arbitrary 
chosen $H,t$-map from (5) and has the result
only if (6) is fulfilled, and in this case
$$
\left\{\aligned \succ (CA3 )\in CA3 [a^{(t)}] ,\ \ &\text{if (7)} ,\\
\succ (CA3 )\notin CA3 [a^{(t)}] ,\ \ &\text{if (8)} .\endaligned\right.
$$

If now we combine the sets of commands of NCA1 , NCA2 and CA3, and put
$\succ (\Cal B )=\succ (CA3),\ E(\Cal B )=E(CA3)$, then in view of Lemma 3 the 
resulting NCA $\Cal B$ will satisfy the conditions of Theorem 1. 

Now let us describe NCA2 and NCA3.
Real work of NCA2 is such
 that at first the group G1 will operate, and after that, 
groups G2 and G3 will operate simultaneously. 
\nn

\subhead NCA2 \endsubhead

 The program contains 3 groups of commands:

G1 : \ Realization of function $H=O(\sqrt{T_{\Cal A} (n)} +S_{\Cal A} (n))$.

G2: \ Isolation of area for modeling.

G3: \ Construction of $A-H-0$-map.
\n

If , for example, $T_{\Cal A} (n) =2^{2n} ,\ S_{\Cal A} (n) =n^4$, then G1 
consists of all commands of the following forms:

$$
\aligned
(c_i ,\dots ,y,?) &\ar \bar c_i ,\\
(\bar c_i ,\dots )&\ar c_i ,\\
(c_i ,\dots ,\bar c_j ,?)&\ar c_i^* ,\\
(c_i^* ,\dots \bar c_j ,?)&\ar \bar c_i ,\\
(d,c_0 ,\dots , \bar c_i )&\ar d' ,
\endaligned \ \ \ \
\aligned
(c_0 ,c_0 ,\dots \bar c_i )&\ar d ,\\
(\bar c_i ,\dots ,c_0 )&\ar c_i^+ ,\\
(c_i^+ ,\dots )&\ar c_i ,\\
(c_i ,\dots ,\neq \bar c_s ,c_j^+ )&\ar c_i^+ ,\\
(y ,\dots ,c_i^+ )&\ar e,
\endaligned
$$
where
$i,j,s$ take all values from $\{ 1,2,\dots \}$, ? means an arbitrary letter, 
$y\in\{ c_0 ,d,d'\}$ , 
$\bar c_i ,c_i^* ,c_i^+ $ are new different special letters for any 
$c_i \in\w$.
\n

 Group G2 consists of all commands of the following forms:
$$
\aligned
(c_0 ,d' ,\dots )&\ar d ,\\
(d,d' ,\dots  )&\ar d'\\
(d' ,\neq d' ,\dots  )&\ar d ,\\
(\dots ,b,\dots )&\ar b,
\endaligned \ \ \ \ 
\aligned
(e ,\dots )&\ar c_0 ,\\
(z_1 ,c_0 ,z_2 ,\dots c_0 )&\ar c_0 ,\\
(z_1 ,c_0 ,c_0 ,\dots ,c_0 )&\ar b ,\\
(c_0 ,c_0 ,e ,\dots )&\ar b ,
\endaligned 
$$
where $z_1 ,z_2 \in\{ d,d'\}$.
\n

Group G3 consists of all commands of the following forms:
$$
\aligned
(c_j ,b ,\dots )&\ar h_j ,\\
(c_0 ,h_i ,\dots )&\ar h_s ,\\
(h_i ,\dots )&\ar h_i ,\\
(h_i ,\dots )&\ar \Gamma _i ,\\
(h_s ,\dots ,\Gamma _j ,\dots ) &\ar \emptyset ,\\
(\Gamma _j ,\dots h_s ,\dots  )&\ar \emptyset ,
\endaligned
$$
where $i,j,s$ take all values from $\{ 1,2,\dots \}$, $h_i$ are special new 
letters, corresponding to each $i$,
and $\Gamma _i$ takes values from the set of all ports $p$ such that 
$con(p)=c_i$.

\n
For $T_{\Cal A} (n) =n^4 ,\ S_{\Cal A} (n) <O(\sqrt{T} )$ the group G1 
consists of all commands of the following forms:
$$
\aligned
(c_0 ,\dots ,c_i )&\ar d ,\\
(c_i ,\dots ,c_0 ,\neq c_j^0 )&\ar c_i^+ ,\\
(c_i^+ ,\dots ,\neq c_0 )&\ar c'_i ,\\
(c_i ,\dots ,c_i^+ ,?)&\ar c_i^* ,\\
(c_i^* ,\dots ,\neq c_0 )&\ar c_i ,\\
(c_i^* ,\dots ,c_0 )&\ar c_i^- ,\\
(c_i ,\dots \neq c_s^+ ,c_j^- )&\ar c_i^- ,\\
(c_i ,\dots c_j^* ,\neq c_s^- )&\ar c_i^* ,
\endaligned \ \ \ \ 
\aligned 
(c_i^- ,\dots ,\neq c'_j ,?)&\ar c_i ,\\
(c_i^- ,\dots ,c'_j ,?)&\ar c_i^+ ,\\
(c'_i ,\dots ,\neq c_j^- )&\ar c'_i ,\\
(c'_i ,\dots ,c_j^- )&\ar c_i ,\\
(c_i^+ ,\dots ,c_0 )&\ar c_i^0 ,\\
(c_i^0 ,\dots )&\ar c_i ,\\
(c_i ,\dots c_j^0 )&\ar c_i^0 ,\\
(z,\dots c_i^0 )&\ar e ,\\
(d,\dots c_i )&\ar d' ,
\endaligned
$$
where $z\in\{ d,d' \}$.
NCA2 is described.
\nn

\subhead CA3 \endsubhead

Put $E(CA3)=\{ c'_k ,c'_{k-1} \}$.
$$
\aligned
(\Gamma _{c_p} ,\neq \emptyset ,b,\undersetbrace{\text{no} \ \emptyset } 
\to {\dots } )&\ar \{ \rho _p ^0 \} ,\\
(\rho ^i_p ,\undersetbrace {\text{no} \ \emptyset }\to {\dots } )&\ar 
\{ \rho _p^{i+1} \} ,\ \ i=0,1,\dots ,2r ,\ p\in\{ k,k-1\} ,\\
(\neq \emptyset ,\neq \emptyset ,\rho _p^{2r+1} ,\undersetbrace{\text{no} \ 
\emptyset }\to {\dots } )&\ar \{\rho _p^0 \} ,\\
(b,\rho _p^i ,\dots )&\ar c'_p ,\\
(\dots ,\emptyset ,\dots )&\ar \emptyset ,
\endaligned
$$
where every letter of NCA1, NCA2 may occur in "$\dots$".
 \n

Theorem 1 is proved.

The general form of "successful" operation of resulting cellular automaton is 
shown on Figure 1.

\n
The technique evolved allows to derive the following amplification of 
Theorem 1.
\n
 
\proclaim{Corollary 1}
If $r_2 >r_1$, then
$$
\NC (r_1 ,T,S)\subseteq\NC (r_2 ,\sqrt{T} +S,(S^{r_1} T^{1/2} )^{1/r_2} +S).
$$ \endproclaim
\n

{\it Proof}

Let $\Cal A$ be a nondeterministic cellular automaton of dimension $r_1$ 
computing predicate $P_{\Cal A}$ with complexity $(T,S),\ S(n)<O(T(n))$. 
Nondeterministic cellular automaton $\Cal B$ of dimension $r_2$ 
simulating $\Cal A$ with complexity 
$(T_1 ,S_1 ),\ T_1 =O(\sqrt{T} +S),\  S_1 =O((S^{r_1} T^{1/2} )^{1/r_2}
 +S)$ can be constructed by a modification
of method from the proof of Theorem 1.

The area for simulation can be organized so that its diameter will be
$$O((S^{r_1} T^{1/2} )^{1/r_2} +S)$$.

For example, for $r_1 =1,\ r_2 =2$ the area from the Figure 1 can be
 constructed 
as a spiral so that the operation of $\Cal B$ will have the form, presented 
on the Figure 2. 
\n 

Here the work of $\Cal B$ in straight strips is the same as in the proof of 
Theorem 1,
and in corner squares $\Cal B$ only verifies the coincidence of the words
written on two sides at each time step, like on Figure 3.
It is not difficult to understand how $\Cal B$ must be arranged, and
I omit awkward details. $\ \square$
\n

\proclaim{Corollary 2} If $\ \ln{T}/\ln{S} =O(1)$, then for all $r=1,2,\dots$
$$
\NC (r,T,S)\subseteq\NC ([\ln{T}/\ln{S} ]+[\log _2 (\ln{T}/\ln{S})]+2+r, S,S).
$$ \endproclaim
\n

{\it Proof}
\n

\proclaim{Lemma 4}
Let $\Cal A$ be NCA, computing $P_{\Cal A}$ with complexity $(T,S),\
S<O(T)$. For every $k=0,1,\dots$ there exists a sequence
$$
\Cal A_0 ,\ \Cal A _1 ,\ \Cal A _2 ,\dots ,\Cal A_k ,
$$
where $\Cal A =\Cal A _0$, for all $i=0,1,\dots \ \ \ \Cal A _i$ is NCA
of dimension $r_i$
and complexity $(T_i ,S_i)$, computing $P_{\Cal A}$,
where for $i>0$
$$
\aligned r_i &\leq r_{i-1} +[\ln{T_i} /2\ln{S} ]+1,\\
T_i &= O(\sqrt{T_{i-1}} +S), \\
S_i &=O(S) .\endaligned
\tag 9$$
\endproclaim

\n
{\it Proof}
\n

Induction on $k$. Basis is evident. Step. Let $k>0$. Applying the inductive 
hypothesis we obtain the sequence $\Cal A_0 ,\ \Cal A _1 ,\ \Cal A _2 ,\dots ,
\Cal A_{k-1}$ with the conditions (9). When $T_{k-1} =O(S_{k-1} )$, we put 
$\Cal A _k =\Cal A _{k-1}$. When $T_{k-1} >O(S_{k-1})$, Corollary 1 with 
$T_{k-1}$ playing the role of $T$ yields the required NCA $\Cal A_k$. Lemma 4 
is proved.

Let $h$ be the least number such that $T_h =O(S)$. In view of 
$T_i = O(\sqrt{T_{i-1}} +S)$ we conclude that 
$h\leq [\log_2(\ln{T}/\ln{S})]+1$. Then it follows from (9) that 
$$
r_h \leq r+h+1+L/2 +L/4 +\dots +L/2^h \leq r+h+1+L ,
$$
where $L=[\ln T /\ln S ]$. Corollary 2 is proved.
\nnn

\specialhead \ \ \ \ \ \ \ \ \ \ \ \ 4. The optimization of 
NCA. Fastest NCA \endspecialhead
\nn
 
{\it Proof of Theorem 2}

\proclaim{Proposition 1} Given a function $\a (n):\ \Bbb N\ar \Bbb N$, 
constructible in time , every $r$-dimensional NCA $\Cal A$ with complexity 
$(T,S)$ can be simulated in $\Bbb Z^r$ with complexity 
$(C_1 T/\a +C_2 S\a ^{1/r} ,\ C_3 S\a ^{1/r} ),\ C_1 ,C_2 ,C_3$ depend on 
$r$.\endproclaim
\n

\subhead Proof\endsubhead
\n

As above, we shall consider the case $r=1$ in more detail. NCA $\Cal A '$, 
simulating $\Cal A$ has the form
$$
\Cal A ' =\Cal A _1 *\Cal A _2 ,
$$
where $\Cal A _1 $ constructs the area for simulating and $\Cal A _2 $ 
simulates $\Cal A$.

Given an input word $B$ , the area for simulating has a form
$$
\dots c_0 bB0^k (b0^p )^\a b_1 c_0 \dots \ ,
\tag 10$$
where $p=k+|S|$. The configuration (10) can be simply constructed in a time 
$O(\a S )$ by $\Cal A _1 $. The configuration (10) is an input one for 
$\Cal A _2$. We suppose that $T/\a =q(n)\in\Bbb N$ and let an evolution
$a^{(1)} ,a^{(2)} ,\dots ,a^{(T)}$ of $\Cal A$ be divided into $\a$ sequential 
segments:
$$
\aligned
\Delta_1 :\ \ &a^{(1)} ,\dots ,a^{(q+1)} ,\\
\Delta_2 :\ \ &a^{(q+1)} ,\dots ,a^{(2q +1)} ,\\
&\dots\\
\Delta_\a :\ \ &a^{(T-q+1)} ,\dots ,a^{(T)} ,
\endaligned
$$
such that input configuration of $\Delta _{j+1}$: $\In (j+1)$ and output 
configuration  of $\Delta _r$ : $\out (j)$ are identical, $j=1,\dots ,\a -1$. 

$\Cal A _2$ simultaneously simulates all possible segments 
$\Delta _1 ,\Delta _2 ,\dots ,\Delta _\a $ at the sites of sequential 
occurrences of the words: $B0^k ,b0^p ,b0^p ,\dots $ in (10). For every 
$\Delta _j$ $\Cal A _2$ stores $\In (j)$ and checks the equality 
$$
\In (j+1)=\out (j)
\tag 11$$
for all $j=1,\dots ,\a -1$. $\Cal A _2$ places a special mark of "flow" in a 
cell where the violating of equality (11) is detected.
At last the letter $b_1$ moves to left through all domains where the 
equality (11) has been verified and if $b_1$ has not met a mark of "flow" 
before $c_0$, then $\Cal A _2$ achieves "success".
All these actions can be fulfilled simultaneously because $\Cal A _2$
is nondeterministic and we obtain that $\Cal A$ achieves "success" 
in time $O(T)$ iff $\Cal A _2$ achieves "success" in time $O(T/\a )$
beginning with (10).

A program of $\Cal A _2$ has the following form:
$$
\aligned
(c,?,?) &\ar (c,c),\\
(0,?,?) &\ar (x,x) ,\\
((c_1 ,c_2 ),(e_1 ,e_2 ),(g_1 ,g_2 )) &\ar (c_1 ,\Cal A (c_2 ,e_2 ,g_2 )),\\
((x,y) ,?,?) &\ar (x^+ ,y^+ ) ,\\
((x^+ ,y^+ ),?,b) &\ar (x^+ ,0 ),\\
(b,(x^+ ,y^+ ),?) &\ar y^+ ,\\
((x^+ ,0 ),y^+ ,?) &\ar (x^+ ,y^- ) ,\\
(y^+ ,?,?) &\ar b,\\
((x^+ ,0),(y^+ ,z^{+,-},?) &\ar (x^+ ,z^{+,-} ) ,\\
((x^+ ,y^{+,-} ),?,(y^+ ,0)) &\ar (x^+ ,0),\\
((x_1^+ ,y^-_1 ),(x_2^+ ,y_2^- ),(x_3^+ ,y_3^- )) &\ar h,\\
\text{where } &\ h=f,\ \text{if}\ x_1 \neq y_1 ,\ \text{or}\ x_2 \neq y_2 ,\ 
\text{or}\ x_3 \neq y_3 ,\\
& \text{else}\ h=s',\\
((x^+ ,y^+ ),u,v) &\ar f,\ \text{where}\ (z,w)\in\{u,v\} ,\\
((x^+ ,y^- ),s' ,b) &\ar ,\\
(b,s'',c_0 ) &\ar s,\\
((x^+ ,y^- )\ \text{or}\ b\ \text{or}\ s'\ \text{or}\ s'' ,?,s) &\ar s,\\
((c,c) ,x,?) &\ar f ,\\
((c,c),?,x) &\ar f,\\
(s,c_0 ,?) &\ar !,
\endaligned
$$
where $\succ(\Cal A _2 )=\ !,\ f$ denotes "flow".
The case $r=1$ is considered. In the cases $r>1$ the input configuration $D_0$ 
for $\Cal A _2$ consists of $\a$ copies of cube $\{ 1,\dots ,S(n)\} ^r$, 
disposed sequentially along a spiral, so that the size of $D_0$ does not 
exceed $O(S\a ^{1/r} )$. Evident changes must be done in the definition of 
$\Cal A _2$. Proposition 1
 is proved. Let us turn to the proof of Theorem 2.
\n

Proposition 1 makes it possible to achieve the time complexity 
$$
T(\a )=C_1 \frac{T}{\a } +C_2 S\a ^{1/r} .
$$

With the aim of finding the minimum of $T(\a )$ we consider the equation:
$$
T' (\a )=-C_1 \frac{T}{\a ^2 } +C_2 S\a ^{\frac{1}{r} -1} /r =0
$$
which yields
$$
\a _{min} =C(T/S)^{\frac{r}{r+1}}
\tag 12
$$
for some constant $C=C(r)$. The function (12) is constructible in time as $T$ 
and $S$. Taking this value of $\a$ for $\Cal A '$ from Proposition 1,
we obtain NCA $\Cal A '$ simulating $\Cal A$ with time and space complexity 
$O(T^{\frac{1}{r+1}} S^{\frac{r}{r+1}} )$. This proves Theorem 2.
\n

\subsubhead Proof of Theorem 3 \endsubsubhead
\n

The assumption that $T>O(S)$ for some fastest NCA contradicts to Theorem 2. 
Theorem 3 is proved.

Hence, while on the subject of fastest NCA we may talk only about their (time) 
complexity $T$, because in view of Theorem 3 $S=O(T)$.
\nnn

\specialhead \ \ \ \ \ \ \ \ \ \ \ \ 5. Complex method of
simulation. Method of evolvents \endspecialhead
\nnn

\subsubhead Proof of Theorem 4\endsubsubhead

\proclaim{Proposition 2} If $r>1$, then $\NC (r,T(n),S(n))\subseteq 
\NC (r-1,T(n)S(n),S^{\frac{r}{r-1}}(n))$.
\endproclaim
\n

\subsubhead Proof \endsubsubhead
\n

Given $r$-dimensional NCA $\Cal A _r$ with complexity $(T,S)$, computing 
predicate $P$, we shall define NCA $\Cal A _{r-1}$ of dimension $r-1$  with 
complexity $(T_{r-1} ,S_{r-1} )$, such that each time step in the evolution 
of $\Cal A$ will be simulated by $\Cal A _{r-1}$ in $O(S(n))$ time steps.

The support of configuration $a^{(t)}$ is the set $S_r^{(t)} =\{ \bar i \in 
\Bbb Z ^r | a(t,\bar i )\neq c_0 \}$.
Without loss of generality we can assume that all supports $S_r^{(t)}$ for 
$\Cal A _r$ are $r$-dimensional cube $B^r ,\ B=\{ 1,\dots ,H\}$, where 
$H=S(|B|)$ depends on the input word $B$. All supports $S_{r-1}^{(\tau )}$ 
of corresponding evolution of $\Cal A _{r-1}$ are contained in 
$r-1$-dimensional cube $B_1^{r-1} ,\ B_1 =\{ 1,\dots ,H_1 \}$ where 
if $p=\min\{ p\in\Bbb N\ |\ pH\geq H^{\frac{r}{r-1}} \}$ then $H_1 =(p+1)H$.

If $V'_x$ denotes a section of $B^r$ by hyperplane $i_r =x ,\ x=1,\dots ,H$, 
then there exist $V_1 ,\dots , V_H \subset B_1^{r-1}$ and fixed isomorphisms 
$V'_x \ar V_x$ such that 
$$
\bigcup\limits_{\tau\in\Bbb N} S_{r-1}^{(\tau )} 
=\bigcup\limits_{x=0}^H V_x \subseteq B_1^{r-1} .
$$

{\bf Definition}. The set $\bigcup\limits_{x=0}^H V_x$ is called an {\it 
evolvent } of $B^r$. An evolution of $\Cal A _{r-1}$ in $V_x$ will simulate 
an evolution of $\Cal A _r$ in $V'_x$.

We shall describe $\Cal A _{r-1}$ in detail only for $r=2$ 
because for $r>2$ $\ \Cal A _{r-1}$ can be constructed along similar lines.
To facilitate further notations we need to introduce some auxiliary notions.
Let $\Cal A$ and $\Cal B$ be one dimensional CA with alphabets $\w_1$ and 
$\w_2$,
determined by programs $\Pi_\Cal A$ and $\Pi_\Cal B$ respectively.

{\bf Definition}. 

1). A {\it composition} of $\Cal A$ and $\Cal B$ is a cellular automaton, 
denoted by $\Cal A *\Cal B$ which is determined by program $\Pi_\Cal A 
\cup\Pi_\Cal B$.

2). A {\it direct product} of $\Cal A$ and $\Cal B$ is a cellular automaton
denoted by $\Cal A\times\Cal B$ with alphabet 
$\w^0 =(\w_1 \times\w_2 )\cup\{ 0(\w_1 )\}$ and program $\Pi$, which consists 
of all commands of the form
$$
((u_1 ,u_2 ),(v_1 ,v_2 ),(w_1 ,w_2 ))\ar (z_1 ,z_2 ) ,
$$
where 
$z_1 =\Cal A (u_1 ,v_1 ,w_1 ),\ z_2 =\Cal B (u_2 ,v_2 ,w_2 ) $ ,
$0(\w^0 ) =0(\w_1 )$.

3). In addition, let $G$ be some set of words of the form:
$$
((\a _1 ,\a _2 ), (\b _1 ,\b _2 ),(\g _1 ,\g _2 ))\ar (\d _1 ,\d _2 ) ,
$$
where $\a_j ,\b_j ,\g_j ,\d_j \in\w_j ,\ j=1,2$.

NCA $\Cal D$ , determined by the program $\Pi\cup G$ is denoted by 
$\Cal A\times _G \Cal B$. It may be called a {\it semidirect product}.

4). Let $\Pi_\Cal A ^*$ be the set of all commands of the form $(x,y,z)\ar\w$, 
where $(x,z,y)\ar\w$ contains in $\Pi_\Cal A$. Then $\Cal A ^*$ denotes NCA 
with program $\Pi_\Cal A ^*$.

5). At last let $\Cal N_\w$ denote a standard automaton in alphabet $\w :\ 
\Cal N_\w (x,y,z)=x$.

\n
\subhead Definition of $\Cal A _{r-1}$ \endsubhead

Put:
$$
\Cal A _{r-1} =\Cal B_{r-1} * \Cal M ,
$$
where $\Cal M$ is a {\it marker of initial evolvent} $a_B$ and $\Cal B_{r-1}$ 
simulates $\Cal A _{r-1} $ in evolvent.

Let $\w =\{ c_0 ,c_1 ,\dots ,c_k \}$ be alphabet of 
$\Cal A _r$, $\w ^+ =\{ c_0^+ ,c_1^+ ,\dots ,c_k^+ \}$ and 
$\w ' =\{ c'_0 ,c'_1 ,\dots ,c'_k \}$ be new different copies of $\w$; 
$0,b$ be new letters. Put $\s =\w\cup\w ^+ \cup\w ' \cup\{ 0,b\}$. 

An auxiliary CA $\Cal W$ in alphabet $\s$ is defined by the following program 
$\Pi$:
$$
\aligned
(c_i ,? ,x) &\ar 0,\\
(0,c_i ,?) &\ar c_i ,\\
(b,c_i ,? ) &\ar c^+_i ,
\endaligned\aligned
(c_i' ,?,0) &\ar 0 ,\\
(0,y,?) &\ar y ,\\
(c_i^+ ,?,?) &\ar b ,
\endaligned
$$
where $x$ takes all values from $\{ 0,b\} ,\ y$ - from 
$\w^+ \cup\w ' ,\ i=0,\dots ,k$.

Let $\w _1 =(\s\times\s\times\s )\cup\{ c_0 \} ,\ G$ be the list of all 
commands of the following form:
$$
 ((u_1 ,u_2 ,u_3 ),(v_1 ,v_2 ,v_3 ),(w_1 ,w_2 ,w_3 ))\ar (z ,z,z ) ,
$$
where $z=\Cal A _r (u_2 ,v_2 ,w_2 ,u_3 ,u_1).$

Put
$$
\aligned 
\Cal B_{r-1} &=(\Cal W\times\Cal N _\s ) \times _G \Cal W ^* ,\\
 \succ (\Cal B_{r-1} ) &=(c_k ,c_k ,c_k ) ,\\
E(\Cal B _{r-1} ) &=\{ (c,c,c)\ | \ c\in E(\Cal A _r )\} ,\\
0(\w _1 ) &= c_0 ,
\endaligned \tag 13
$$
where $c_k =\succ \Cal A _r ,\ c_0 =0(\w )$.

Given an input word $B\in\w ^*$ for $\Cal A _r$,
let $d(r,s)$ be the initial configuration of $\Cal A_r$, corresponding to $B$, 
we define corresponding input word $a_B =h_0 h_1 \dots h_{H^2} ,\ 
H=S_{\Cal A_r} (|B|)$ for $\Cal B _{r-1}$ by the following:
$$
h_i =\left\{\aligned
(b,b,b) ,\ &\text{if}\ i\equiv 0 \ (\mod H) ,\\
(c_0 ,c_0 ,c_0 ) ,\ &\text{if}\ i\equiv r\ (\mod H),\ q<r<H,\\
d(r,s) ,\ &\text{if}\ i=sH+r,\ 1\leq r\leq q.
\endaligned\right.
$$

The following Lemma 5 may be deduced immediately from the definition (13).

\proclaim{Lemma 5} $\forall B\in P_{\Cal A _r}\ \ \tau_{\Cal B _{r-1}} 
(a_B )\leq 2H\tau_{\Cal A_r} (B) .$\endproclaim
\n

Now to finish the proof of Proposition 2 it is sufficient to construct CA 
$\Cal M$ transforming $B$ to $a_B$ in time $O(H^2 )=O(H^{\frac{r}{r-1}} )$. 
It may be done as in the proof of Theorem 1 (look at NCA 2: G1 and G2). Note 
that here we have $V_x =\{ (x-1)H+1 ,\dots ,xH\}$.

The case $r=2$ is considered.

Let now $r>2$. This case differs in that the domains $V_x$ for sections 
$i_r =x ,\ x=0,1,\dots ,H ,\ H=O(S)$ are disposed sequentially along a spiral 
so that $S_{r-1} =O(S^{\frac{r}{r-1}} ),\ T_{r-1} =O(TS)$. Such initial 
evolvent
$a_R \subset \Bbb Z ^{r-1}$ can be isolated from the space and marked out 
according to the direction of laying of all $V_x$ into evolvent in time
$O(S)$.

 Proposition 2 is proved.

Note that analogous proposition takes place also for deterministic CA.

\nn
Now we turn to the proof of Theorem 4.

1). Given a fastest $r$-dimensional NCA $\Cal A$ with complexity $T=T_{r,P} ,\ 
P=P_\Cal A $, Proposition 2 yields $r-1$-dimensional NCA $\Cal A '$ simulating
 $\Cal A $ with complexity $(T^2 ,T^{\frac{r}{r-1}} )$. Then by Theorem 2
 we obtain $r-1$-dimensional NCA simulating $\Cal A '$ with complexity 
$O( T^{2/r} (T^{\frac{r}{r-1}} )^{\frac{r-1}{r}} )=O(T^{1+2/r } )$. 
Point 1) is proved.

\n
2). We can suppose that $\Cal A$ acts in $r$-dimensional cube $B^r ,\ 
B=\{ 1,\dots , S\}$. $r+1$-dimensional NCA simulating $\Cal A$ in time 
$O(T^{1-r/(r+1)^2} )$ will be constructed in two steps.

Step 1. Simulation of $\Cal A$ in $\Bbb Z^{r+1}$ with complexity 
$(T,T^{\frac{r}{r+1}} )$.

Step 2. Applying of Theorem 2 to NCA obtained in Step 1.
\n

{\bf Step 1}.

{\bf Definition}. A set $A$ consisting of inclusions of the form
$$
L_j :\ Y_j \ar \Bbb Z^r ,\ j\in\Bbb N ,
$$
is called  a constructible set of inclusions if 
for some $q \leq r$ all $Y_j \subset \Bbb Z^q$ , and
 for some constants $c,c_1 ,c_2$ the following three conditions are 
satisfied.

a). $\ \forall j\in\Bbb N ,\ \bar x,\bar y\in Y_j$
$$
\rho (L_j (\bar x) ,L_j (\bar y) )\leq c\rho (\bar x,\bar y),
\tag 14$$
where $\rho$ denotes standard metric in $\Bbb Z^q$ or $\Bbb Z^r$.

b). $\ \Diam (\Im L_j )\leq c_1 |L_j |^{1/r}$, where $|L|$ denotes the number 
of elements in $L$.

c). All domains $\Im L_j$ can be marked out by NCA in time $c_2 |L_j |^{1/r}$ 
so that every cell $\bar z=L_j (\bar x)\in\Im L_j$ will be marked by a label 
$m(\bar z)$ which points to the disposition of all such cells $L_j (\bar y)$ 
that $\rho (\bar x,\bar y)=1$ with respect to $\bar z$. (Every label 
$m(\bar z)$ has the form $\langle \bar g_0 ,\bar g_1 ,\dots ,\bar g_{2q} 
\rangle$ where $\bar g_p =L_j (\bar x (p))-\bar z\in\Bbb Z^r ,
\ p=0,1,\dots ,2q;\ \bar x (p)$ is defined in the section 1. In view of 
inequality (14) the required number of all such labels $m(\bar z)$ does 
not depend on $j$.)

\nn
To fulfill Step 1 it is sufficient to prove the following

\proclaim{Proposition 3} For every $r$ there exists a constructible set of 
inclusions $L_S^r :\ B^r \ar \Bbb Z^{r+1} ,\ 
B=\{ 1,\dots ,S\} ,\ S=1,2,\dots $.
\endproclaim
\n

Really, with such inclusions $L_S^r$ in view of the point b) we can simulate 
$\Cal A$ with complexity $(T,O(S^{\frac{r}{r+1}} ))$ in $\Bbb Z^{r+1}$.

\proclaim{Lemma 6} Given $r\leq 1$ there exists a constructible set of 
inclusions
$$
R_S^r :\ Y_s \ar \Bbb Z ^{r+1} ,\ S=1,2,\dots ,
$$
where $Y_S =B_1^r \times B \in\Bbb Z^{r+1} ,\ B_1 =\{ 1,\dots ,S_1 \} ,\ S_1 
=[S^{\frac{r-1}{r}} ]$.
\endproclaim
\n

\subhead Proof \endsubhead
\n

A parallelepiped $Y_S$ in $\Bbb Z^{r+1}$ can be conceived of as a thread 
of length $S$ and $S^{\frac{r-1}{r}}$ thick. This thread can be rolled up 
into a ball which has the shape of cube of side $4|Y_S |^{\frac{1}{r+1}}$ 
in $\Bbb Z^{r+1}$, because 
$S^{\frac{r-1}{r}} <|Y_S |^{\frac{1}{r+1}} =S^{\frac{r}{r+1}}$.
The construction of NCA performing the required marking of $\Im Y_S$ is evident. $\square$
\n

\subhead Proof of Proposition 3 \endsubhead
\n

Induction on $r$. Basis $r=1$. Proposition 3 follows from Lemma 6. Step: 
$r>1$.  It follows from the inductive hypothesis that there exists a 
constructible set of inclusions:
$$
L_S^{r-1} :\ B^{r-1} \ar B_1^r ,\ S=1,2,\dots .
$$

Let $R_S^r$ be inclusions from Lemma 6. Then we obtain a constructible set 
of inclusions $ L_S^r :\ B^r \ar \Bbb Z^{r+1}$, defined by the following: 
$$
L_S^r (\bar x ,y)=R_S^r (L_S^{r-1} (\bar x ),y),
$$
 where $\bar x\in\Bbb Z^{r-1},\ y\in\Bbb Z$.

$\square$

Step 1 is fulfilled.
\n

{\bf Step 2}.

Applying Theorem 3 we obtain a simulation of $\Cal A$ in $\Bbb Z^{r+1}$ with 
complexity 
$T^{\frac{1}{r+1}} (T^{\frac{r}{r+1}} )^{\frac{r}{r+1}} =T^{1-r/(r+1)^2}$.

Theorem 4 is proved.
\n

Thus, if $T_\Cal A >S_\Cal A$, then we have a variety of ways to accelerate 
the computations of $P_\Cal A$ in $r+1$-dimensional space, for example:

1). By Theorem 1 (or by Corollary 1).

2). Sequential applications of Theorem 2 and point 2) of Theorem 4.

The second way gives the better acceleration for $T_\Cal A \gg S_\Cal A$.
But if $T$ is small, for example $T=S^2$, then Theorem 1 gives the more strong 
estimate for $T_{r+1 ,P}$ because $1<\frac{r+2}{r+1} (1-\frac{r}{(r+1)^2} )$.

\nn

\proclaim{Remark} For any predicate $P$ computible in time $T=O(n^\a ),\ \a 
=const$ on $d$-dimensional NCA and for any $\b >0$ there exists such a number 
$r\geq d$ that
$T_{r,P} =O(n^\b )$.
\endproclaim
\n

\subhead Proof \endsubhead
\n

Applying Theorem 4, point 2), we obtain that for $r>d$
$$
T_{r,P} (n)\leq c(R) \exp [\ln (T_{d,P} (n)) 
\prod\limits_{m=d+1}^r (1-m/(m+1)^2 )].
$$
We have:
$$
\ln \prod\limits_{m=d+1}^r (1-m/((m+1)^2 )\sim -\sum\limits_{m=d+1}^r 
m/(m+1)^2 \ar -\infty\ (r\ar +\infty ).
$$
Thus, $\prod\limits_{m=d+1}^r (1-m/(m+1)^2 )\ar 0\ (r\ar\infty ). \ \square$

\nn
\specialhead\ \ \ \ \ \ \ \ \ \ \ \ \ 6. Discussion\endspecialhead
\nn

We see that if cellular automata are nondeterministic, then increase of 
dimension
leads to the substantial acceleration of computations.
In addition, programming on nondeterministic CA is more simple, then on 
ordinary CA.
Thus, if it is possible, the realization of many dimensional nondeterministic 
cellular automata by a physical device would 
be of great practical consequence. From the other side, TCD- problem
for deterministic CA remains unsolved.
 Let $C(r,T,S)$ be the class of 
predicates, computable on CA in time $T$ and space $S$. TCD-problem 
for CA is as follows: 

{\it Given $r,S,T$, is there an increasing function $f(n)$ such that 
$$
C(r,T,S)\subseteq C(r+1,\frac{T}{f(n)} +S ,S_1 )
$$
for some $S_1$?}

At last note that the hypothesis $T=O(f(S))$ for the fastest deterministic CA 
is open for question for every function $f(n)\geq n$ growing slowly in 
comparison with exponential.

\nnn
\specialhead \ \ \ \ \ \ \ \ \ \ \ \ 7. Acknowledgements \endspecialhead
\nn

I am grateful to Alexander Shen for useful information on the theory of 
complexity, and Nadia Viktorova for editorial help. 
\nn

\enddocument
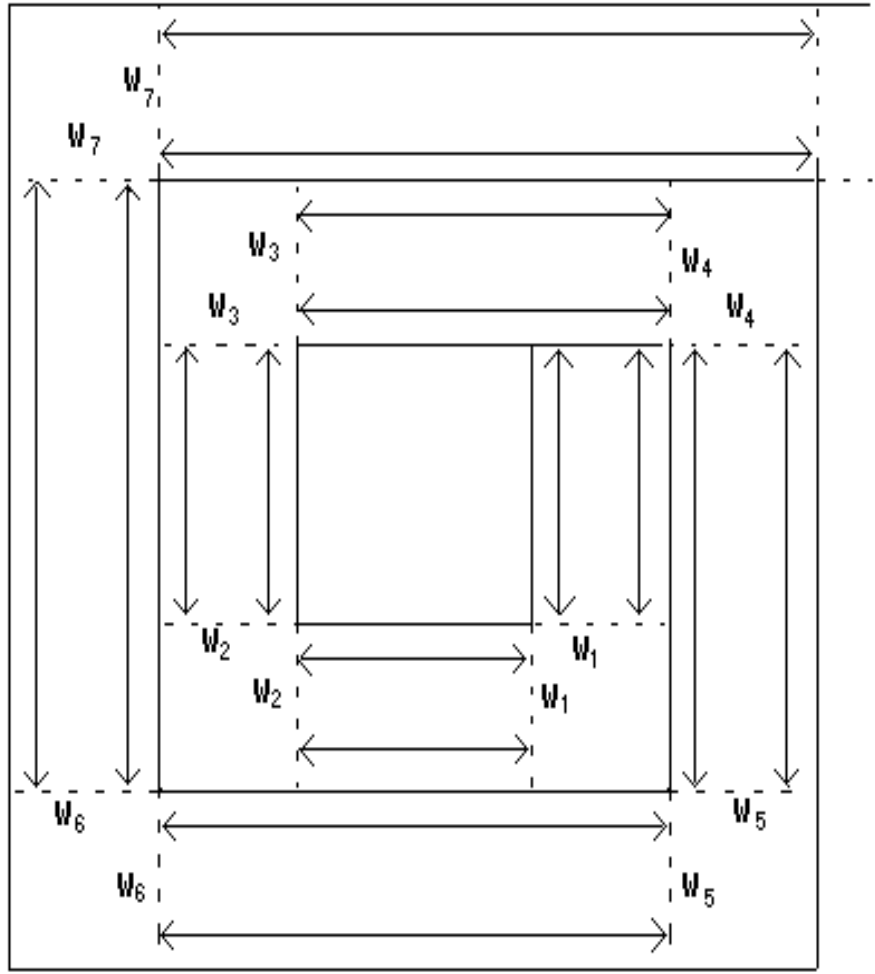
